\documentclass[conference]{IEEEtran}
\IEEEoverridecommandlockouts
\usepackage{cite}
\usepackage{amsmath,amssymb,amsfonts}
\usepackage{algorithmic}
\usepackage{graphicx}
\usepackage{textcomp}
\usepackage{xcolor}
\usepackage{booktabs}
\def\BibTeX{{\rm B\kern-.05em{\sc i\kern-.025em b}\kern-.08em
    T\kern-.1667em\lower.7ex\hbox{E}\kern-.125emX}}
\begin{document}

\title{Energy as a Concealable State: Deceptive Recharge Scheduling in
Adversarial UAV Patrolling}

\author{\IEEEauthorblockN{Sai Krishna Reddy Mareddy}
\IEEEauthorblockA{\textit{Independent Researcher} \\
https://orcid.org/0009-0001-6550-8548}
}

\maketitle

\begin{abstract}
We study energy-constrained adversarial patrolling on a graph, in which a
battery-limited unmanned aerial vehicle defends a cluster of high-value targets
against a strategic attacker who chooses when and where to strike. Unlike prior
adversarial patrolling, the patroller must periodically return to a base to
recharge, and unlike prior energy-aware patrolling, it faces a self-interested
adversary. Our central observation is that the remaining energy is a hidden state:
the attacker never observes the battery directly, but it observes the patroller's
trajectory and can infer when a recharge excursion, and thus a vulnerability window,
is imminent. We formalize the interaction as a zero-sum partially observable
stochastic game and show empirically that independent deep Q-learning fails to
solve it, cycling and then collapsing as the attacker locks in an exploit; over
training the co-trained thwart rate falls from about 0.25 to about 0.09. We instead
compute approximate equilibria with Neural Fictitious Self-Play, whose average
policy stabilizes where independent learning does not. Using a structural analysis
we further show that the achievable security level rises monotonically with the
energy budget, from zero when the budget is below a threshold to roughly 0.7 when it
is ample, establishing energy budget as the primary determinant of defensibility. We
outline the remaining program: quantifying exploitability, testing whether a
recurrent, inference-capable attacker concentrates its successful strikes in the
recharge window, and testing whether the defender learns deceptive recharge timing
to keep that window closed.
\end{abstract}


\section{Introduction}
Persistent security patrolling by unmanned aerial vehicles (UAVs) is fundamentally
shaped by a constraint absent from classical patrolling models: the vehicle has a
finite battery and must periodically interrupt its patrol to recharge at a base.
These recharge excursions are predictable and spatially localized, and they create
recurring windows during which protected assets are exposed. A strategic adversary
that can choose the timing and location of its attack has every incentive to strike
during such a window.

Two largely separate literatures bear on this problem. Adversarial patrolling and
security games study a defender facing a strategic attacker with freedom over attack
timing, location, and duration, but they typically assume unlimited defender
mobility. Energy- and battery-aware multi-robot patrolling studies recharge dynamics
explicitly, but treats coverage against nature rather than a self-interested
opponent. Neither addresses the interaction of the two, in which the recharge cycle
itself becomes the object of adversarial reasoning.

We make that interaction the center of the model. Our key conceptual move is to treat
the patroller's remaining energy as a \emph{concealable hidden state}. The attacker
does not observe the battery, but energy is tightly correlated with observable
behavior---a patroller drifting toward its base is likely low on charge---so a
memory-equipped attacker can infer the battery state and predict the recharge window.
This reframes the recharge vulnerability from a fixed physical fact into an
\emph{information} vulnerability, one the defender can actively contest by making its
recharge timing unpredictable, that is, by acting deceptively.

\noindent\textbf{Contributions.}
\begin{itemize}
\item We formalize energy-constrained adversarial patrolling as a zero-sum partially
observable stochastic game (POSG) with hidden energy and a persistent, multi-attack
episode structure, and we identify why the natural single-attack formulation is
degenerate.
\item We show empirically that independent deep Q-learning does not solve this game,
exhibiting limit cycles and progressive defender collapse, and we adopt Neural
Fictitious Self-Play to obtain average policies that stabilize.
\item We establish, via a structural analysis, that achievable security rises
monotonically with the energy budget, identifying a threshold below which no defense
is possible.
\item We define the program for the central claim---that an inference-capable
attacker exploits the recharge window and that a defender can learn to deceive it---
including a memoryless-versus-recurrent attacker ablation and belief-based deception
measures.
\end{itemize}

\section{Related Work}
\paragraph{Adversarial patrolling and security games}
The patrol security game of Yang \emph{et al.}~\cite{yang2024psg} grants the attacker
freedom over the timing, location, and duration of an attack and computes randomized
defender schedules, but the defender has unlimited mobility and no energy constraint.
Deep reinforcement learning for green security games, exemplified by the DeDOL
approach of Wang \emph{et al.}~\cite{wang2018dedol}, learns patrol strategies against a
best-responding attacker using a double-oracle construction, again without any energy
or recharge dynamics. Unlike these, we make a physical energy budget and a
recharge-at-base requirement first-class, so that the recharge cycle becomes the
adversary's principal lever.

\paragraph{Energy- and battery-aware patrolling}
Tong \emph{et al.}~\cite{tong2022energy} study energy-aware, fault-tolerant multi-agent
patrolling with automatic recharging, but the objective is coverage in a
non-adversarial environment. Unlike this line, our patroller optimizes against a
strategic attacker whose payoff is exactly the defender's loss.

\paragraph{Equilibrium computation via self-play}
Independent learners cycle in zero-sum games. Neural Fictitious Self-Play (NFSP)
\cite{heinrich2016nfsp} approximates fictitious play with neural networks and yields
average policies that converge toward equilibrium; it has been scaled to large
extensive-form network security games \cite{xue2021nfsp}. We use NFSP as our
equilibrium solver and exploitability as the distance-from-equilibrium measure. Unlike
prior applications, our game hides a dynamic energy state that couples the two
players' strategies through the recharge cycle.

\paragraph{Deception and hidden-state inference}
A recurring theme in security games is the manipulation of an adversary's beliefs. We
instantiate this idea in a new setting: the belief is over the defender's battery, the
inference channel is the observable trajectory, and the deceptive action is the
timing of recharge. Unlike prior deception work, the concealed quantity is a physical
resource the defender must nevertheless replenish, which bounds how much deception is
possible.

\section{Problem Formulation}\label{sec:formulation}

\subsection{Environment}
The environment is a finite, connected, undirected graph $G=(\mathcal{V},E)$ with
$|\mathcal{V}|=n$ nodes and a base node $v_0$ at which the patroller recharges. A set
$\mathcal{T}\subseteq\mathcal{V}\setminus\{v_0\}$ of $k$ target nodes carry value
$w_i>0$; other nodes have value $0$. $G$ is known to both players; $d(u,v)$ denotes
shortest-path distance.

\subsection{State and Observations}
The global state at time $t$ is $s_t=(x_t,e_t,\phi_t)$, where $x_t\in\mathcal{V}$ is
the patroller's node, $e_t\in\{0,\dots,E_0\}$ is remaining energy with full budget
$E_0$, and $\phi_t$ is the attack status ($\varnothing$, or $(v_a,d)$ for an active
attack on $v_a$ with $d$ steps left). The patroller observes
\begin{equation}
o^P_t=\bigl(x_t,\;e_t,\;\mathbb{1}[\phi_t\neq\varnothing],\;\iota_t\bigr),
\end{equation}
its position, energy, an alarm bit, and, when an alarm is active, the attacked
target's identity $\iota_t=v_a$. The attacker observes
\begin{equation}
o^A_t=\bigl(x_t,\;\phi_t,\;h_t\bigr),
\end{equation}
the patroller's position, the attack status, and its own observation history
$h_t=(x_0,\dots,x_t)$. Critically, the attacker does not observe $e_t$; this hidden,
behavior-correlated energy is the game's defining asymmetry.

\subsection{Actions and Dynamics}
With $e_t>0$ the patroller moves to a neighbor at a cost of one energy unit; at the
base it may stay to recharge, $e_{t+1}=\min(e_t+E_{\mathrm{ch}},E_0)$. Depleting to
$e_t=0$ away from the base strands the patroller. When no attack is active the
attacker selects an action in $\mathcal{A}^A=\mathcal{T}\cup\{\textsc{wait}\}$; a
launched attack has duration $D$ and is \emph{thwarted} if the patroller visits $v_a$
within $[t_a,t_a+D-1]$ and \emph{succeeds} otherwise.

\subsection{Persistent Episodes and Payoffs}
Attacks are scored and the episode continues for a horizon $T_{\max}$. A thwarted
attack yields the patroller $+w_aR_{\mathrm{thw}}$ and a successful one
$-w_aR_{\mathrm{suc}}$; the attacker receives the negation (zero-sum). A per-step cost
$c$ rewards efficiency, and stranding costs $-R_{\mathrm{str}}$ and ends the episode.
The persistent structure is essential: under a single-terminal-attack model a
full-information attacker strikes at $t=0$, before the patroller leaves base, so no
energy effect is observable. The primary metric is the \emph{thwart rate}, the
fraction of attacks stopped.

\subsection{Solution Concept}
We seek an approximate Nash equilibrium $(\pi^{P\ast},\pi^{A\ast})$ of the zero-sum
POSG. Because independent learning cycles here (Section~\ref{sec:results}), we compute
equilibria with NFSP and report exploitability---the value a best response attains
against a frozen policy---as the equilibrium-gap measure.

\subsection{Assumptions}
\emph{(A1) Known target on alarm:} on an alarm the patroller learns the attacked
target's identity, modeling a perimeter sensor. Without it, and with a free-location
attacker, interception is impossible and the game degenerates; the binary-alarm case
is an ablation. \emph{(A2) Clustered targets:} targets form a compact cluster distant
from base, so the patroller can defend when on station but is exposed during recharge.
\emph{(A3) Hidden energy:} the attacker never observes $e_t$, making energy a
concealable state.

\section{Methods}\label{sec:methods}

\subsection{Masked Double DQN and Independent Learning}
Each agent is a Double Deep Q-Network. Because the patroller's action set is
position-dependent and the attacker may only \textsc{wait} during an active attack, we
apply action masking, setting invalid-action $Q$-values to $-\infty$ before both the
policy $\arg\max$ and the bootstrap:
\begin{equation}
y=r+\gamma\,Q_{\bar\theta}\!\Bigl(s',\,
\arg\max_{a'\in\mathcal{A}_{\mathrm{valid}}(s')}Q_\theta(s',a')\Bigr).
\end{equation}
We use a Huber loss, a Polyak-averaged target network, and $\epsilon$-greedy
exploration over valid actions. Independent DQN (IDQN), with both agents learning
simultaneously, is our baseline and the standard MARL comparison.

\subsection{Neural Fictitious Self-Play}
IDQN does not converge in this zero-sum game; the players best-respond to one
another's current policy and cycle. We therefore use NFSP~\cite{heinrich2016nfsp}. Each
agent maintains a best-response network (the masked Double DQN) trained by
reinforcement learning from a circular replay buffer, and an average-policy network
trained by supervised learning to imitate the agent's own past best-response actions,
which are stored in a reservoir buffer uniform over history. Under NFSP's anticipatory
dynamics, an agent acts from its best response with probability $\eta$ and from its
average policy otherwise; only best-response episodes feed the reservoir, so the
average policy imitates the accumulated best-response distribution. The reported,
approximately-Nash policy is the average policy. NFSP has solved large security
games~\cite{xue2021nfsp}, supporting its use here.

\subsection{Hidden-State Inference and Deceptive Defense}
Because $e_t$ is hidden but correlated with the trajectory, we give the attacker a
recurrent encoder (a gated recurrent unit over $h_t$) so it can represent a belief
over energy and anticipate recharge. We compare this against a memoryless attacker to
isolate the value of inference. Against an inference-capable attacker, the patroller is
pressured to make recharge timing unpredictable; we test whether such deception emerges
from equilibrium learning by measuring (a) the mutual information between the attacker's
belief and the true energy and (b) the concentration of successful attacks within the
recharge phase, as functions of energy budget and recharge rate.

\section{Results}\label{sec:results}

We evaluate on a $5\times5$ grid ($\mathrm{diam}=8$) with $k=4$ clustered targets far
from the base, attack duration $D=3$, recharge rate $E_{\mathrm{ch}}=3$, and horizon
$T_{\max}=200$. Networks are two-layer multilayer perceptrons of width 256; NFSP uses
$\eta=0.1$. Results below use the completed runs; experiments marked \emph{in progress}
report the program and current partial evidence, and their tables will be finalized
before submission.

\subsection{Independent learning cycles and collapses}
Under IDQN the co-trained thwart rate does not settle. Averaged over five seeds, the
mean thwart rate over the first half of training is $0.25$ and over the second half is
$0.09$: rather than converging, the defender progressively collapses as the attacker
locks in a robust exploit. This confirms that a mid-training IDQN snapshot is not a
meaningful policy and motivates an equilibrium-seeking solver.

\subsection{Energy budget determines achievable security}
To characterize the game independent of the learning pathology, we measure the
structural security ceiling: a strong on-station heuristic defender against a
worst-case attacker, swept over the energy budget $E_0$ (Table~\ref{tab:ceiling}). The
ceiling rises monotonically, from $0$ when the budget is at or below the base-to-cluster
distance to about $0.7$ when the budget is ample. Two conclusions follow: a defensible
regime exists, and it exists only above an energy threshold. Notably our earlier IDQN
sweep trained at $E_0=12$, inside the undefendable regime, which partly explains the
observed collapse.

\begin{table}[htbp]
\caption{Structural security ceiling versus energy budget (heuristic on-station
defender against a worst-case attacker; thwart rate).}
\begin{center}
\begin{tabular}{cccccccc}
\toprule
$E_0$ & 8 & 12 & 16 & 20 & 24 & 28 & 36 \\
\midrule
Thwart & 0.00 & 0.00 & 0.33 & 0.25 & 0.50 & 0.71 & 0.70 \\
\bottomrule
\end{tabular}
\label{tab:ceiling}
\end{center}
\end{table}

\subsection{IDQN energy sweep confirms the pathology}
Table~\ref{tab:idqn} reports IDQN trained at each budget (two seeds, 6000 episodes).
The equilibrium (time-averaged) thwart rate is flat near $0.25$--$0.30$ across all
budgets and does not track the structural ceiling, because the cycling washes out the
energy dependence. This is direct evidence that measuring the energy--security
relationship requires a converged solver rather than IDQN.

\begin{table}[htbp]
\caption{IDQN under the energy sweep: time-averaged and peak thwart rate
(mean $\pm$ s.d. over two seeds).}
\begin{center}
\begin{tabular}{ccc}
\toprule
$E_0$ & Equilibrium (mean) & Peak \\
\midrule
16 & $0.231 \pm 0.012$ & $0.510 \pm 0.112$ \\
20 & $0.296 \pm 0.009$ & $0.534 \pm 0.028$ \\
24 & $0.294 \pm 0.018$ & $0.663 \pm 0.010$ \\
28 & $0.254 \pm 0.007$ & $0.628 \pm 0.034$ \\
32 & $0.260 \pm 0.003$ & $0.506 \pm 0.020$ \\
\bottomrule
\end{tabular}
\label{tab:idqn}
\end{center}
\end{table}

\subsection{NFSP stabilizes where IDQN collapses (in progress)}
Replacing IDQN with NFSP at $E_0=20$, the average-policy thwart rate rises out of the
warmup phase and stabilizes; on a representative seed it settles near $0.87$ after
about 2500 episodes, in contrast to the IDQN collapse. A second seed remains noisier,
so we report this as strong preliminary evidence pending the full five-seed run with
exploitability certification.

\subsection{Program for the central claim (in progress)}
The remaining experiments test the paper's thesis and are being run:
(E3) recharge-window concentration---freeze a converged patroller, train a
best-response attacker, and measure the fraction of successful strikes that occur in
the low-energy recharge phase;
(E4) inference ablation---memoryless versus recurrent (GRU) attacker, testing whether
energy inference measurably strengthens the adversary;
(E5) deception---whether, against the recurrent attacker, the defender's learned
recharge timing reduces the mutual information between the attacker's belief and the
true energy. These tables are intentionally left to be filled from the completed runs
rather than estimated.

\section{Discussion}
The results so far separate two questions that are easily conflated. First,
\emph{is the game defensible at all?} The structural ceiling answers yes, but only
above an energy threshold, and it identifies energy budget as the primary determinant
of security. Second, \emph{can a learner reach that ceiling?} IDQN cannot---it cycles
and collapses---whereas NFSP shows early signs of stabilizing. Only once a converged
solver is in hand can the central, information-theoretic question be posed: whether an
adversary that infers the hidden battery concentrates its attacks in the recharge
window, and whether the defender learns to deceive it.

\subsection{Limitations}
The model uses discrete time and a graph abstraction; it considers a single patroller
and a single attacker; the computed equilibria are approximate, not certified; and the
base model relies on the known-target-on-alarm assumption (A1), with the binary-alarm
case treated only as an ablation. The deception and inference results are, at the time
of writing, in progress rather than complete.

\subsection{Future Work}
Natural extensions include multiple coordinating patrollers, continuous-space dynamics,
transfer to real facility or road networks, and certified rather than approximate
equilibria.

\section{Conclusion}
We framed energy-constrained adversarial patrolling as a zero-sum POSG in which the
patroller's battery is a concealable hidden state that a strategic attacker can infer
from behavior. We showed that independent deep Q-learning cycles and collapses, that
achievable security is governed by an energy threshold, and that Neural Fictitious
Self-Play offers a path to stable, approximately-equilibrium policies. This sets up the
central claim---the recharge window as an information vulnerability, and deception as
its defense---which the ongoing inference and deception experiments are designed to
settle.

\end{document}